\documentclass[twocolumn,preprintnumbers,groupaddress,footnote]{revtex4-1}

\pdfoutput=1

\usepackage{amsmath}
\usepackage{graphics}
\usepackage{graphicx}
\usepackage[mathcal]{eucal}
\usepackage{color}
\usepackage{appendix}

\newcommand{\Feyn}[1]{#1\kern-0.45em/}

\begin{document}
\preprint{UCI-TR-2011-27}

\title{Dirac Leptogenesis with a Non-anomalous $U(1)^{\prime}$ Family Symmetry} 

\author{Mu-Chun Chen}
\email[]{muchunc@uci.edu}
\author{Jinrui Huang}
\email[]{jinruih@uci.edu}
\author{William Shepherd}
\email[]{shepherd.william@uci.edu}
\affiliation{Department of Physics and Astronomy, University of California, Irvine, California 92697-4575, USA}

\date{\today}

\begin{abstract}
\label{abstract}
We propose a model for Dirac leptogenesis based on a non-anomalous $U(1)^{\prime}$ gauged family symmetry. The anomaly cancellation conditions are satisfied with no new chiral fermions other than the three right-handed neutrinos, giving rise to stringent constraints among the charges. Realistic masses and mixing angles are obtained for all fermions. The model predicts neutrinos of the Dirac type with naturally suppressed masses. Dirac leptogenesis is achieved through the decay of the flavon fields. The cascade decays of the vector-like heavy fermions in the Froggatt-Nielsen mechanism play a crucial role in the separation of the primodial lepton numbers. We find that a large region of parameter space of the model gives rise to a sufficient cosmological baryon number asymmetry through Dirac leptogenesis.
\end{abstract}


\maketitle

\section{Introduction}
\label{sec:intro}

The origin of the observed baryon number asymmetry of the universe (BAU)~\cite{ref:KolbTurner,ref:barAsyExp} has been a long standing puzzle in particle physics and cosmology. While the Stadard Model (SM) satisfies the three neccessary Sakhrov conditions~\cite{Sakharov:1967dj} to dynamically generate the BAU, there exist several difficulties. The possibility of generating the appropriate BAU using only the known baryon-number violating sphaleron dynamics of the SM $SU(2)_L$ gauge group has been investigated, but requires the Higgs boson to be lighter than is allowed by experimental searches to attain the out-of-equilibrium condition through a strong first order phase transition (for reviews, see {\it e.g.}~\cite{Riotto:1998bt,Riotto:1999yt}). In addition, the amount of CP violation in the CKM matrix leads to highly suppressed asymmetry that is not sufficient to explain the observation~\cite{Shaposhnikov:1986me}.   

Many attempts based on physics beyond the SM have been made in the context of GUT models by producing the BAU at the GUT scale through decays of the heavy vector bosons of the GUT group. More recently, it was noted that it is possible to explain the BAU by generating a lepton-number asymmetry in the early universe and then allowing the $SU(2)_L$ sphalerons to convert that asymmetry partially into baryon number (for reviews, see, {\it e.g.}~\cite{Buchmuller:2005eh,Davidson:2008bu,ref:ChenTASI2006}), thus generating the appropriate BAU without introducing new baryon-number violating processes which are tightly constrained by proton decay searches. 

This leptogenesis framework has been explored in detail in connection with the need for new physics beyond the SM to explain the observation that neutrinos are massive. Most commonly, leptogenesis is achieved by positing heavy Majorana right-handed neutrinos which naturally lead to the appropriate masses for their left-handed partners through the seesaw mechanism, and then allowing them to decay with a CP-violating phase, thus producing a net lepton number asymmetry~\cite{Buchmuller:2005eh,Davidson:2008bu,ref:ChenTASI2006}. Thus the dynamics which lead to the lightness of the neutrinos also generate the asymmetry which can explain the BAU.

It was first pointed out in Ref.~\cite{ref:dirLep1} that if an asymmetry can be built up
between left- and right-handed leptons, and if that asymmetry can survive until
the sphalerons decouple, the BAU could be explained without introducing any
new violation of total lepton or baryon number. This is because the sphaleron dynamics  
do not actually couple to the total lepton number, but only to
left-handed lepton number. 
%This reopens the possibility that $B-L$is a fundamental symmetry of the universe. 
%This requires 
This implies that leptogenesis is possible even if neutrinos are Dirac particles, and thus the mechanism is known as Dirac leptogenesis. 

While there has been tremendous progress over the past decade in understanding the properties of neutrinos, there are still many open questions, including the Dirac versus Majorana nature of the neutrinos. For Majorana neutrinos, seesaw mechanisms provide a natural explanation of the smallness of their masses. For Dirac neutrinos, however, it is much more difficult to realize small neutrino masses. In the context of Dirac leptogenesis, many realistic models have been illustrated in Ref.~\cite{ref:dirLep2}, in which the smallness of the neutrino masses are generated through different mechanisms. Here we take a different approach. By introducing an additional $U(1)^{\prime}$ family symmetry and with the appropriate $U(1)^{\prime}$ charge assignment for the fermions, the usual dimension-4 operators for Dirac neutrino masses can be forbidden. Dirac neutrino mass terms can arise only through the higher dimensional operators and thus are highly suppressed. This is an application of the Froggatt-Nielson mechanism~\cite{ref:frogNiel} which can also give rise to a realistic fermion mass hierarchy and mixing pattern.

Our goal in this work is to reconnect the dynamics responsible for explaining the smallness of neutrino masses with those responsible for the BAU in the Dirac leptogenesis framework. To do so we first introduce a model of flavor for the SM quarks and leptons, based on the Froggatt-Nielsen mechanism. We then study the implications of the UV-completion of the model for Dirac leptogenesis. In particular, the exotic fields, {\it i.e.} the flavon fields and heavy vector-like fermions, required in the Froggatt-Nielsen mechanism play an important role in generating the primordial asymmetry. A distinguishing feature of our model with respect to other Dirac leptogenesis models~\cite{ref:dirLep2} is that the separation of left-handed lepton number asymmetry from the right-handed lepton number asymmetry is through the cascade decays of the heavy vector-like fermions. 

This paper is organized as follows: in Section~\ref{sec:model} we present a flavor model based on a non-anomalous $U(1)^{\prime}$ family symmetry. In Section~\ref{sec:diracLep} we show how the Dirac leptogenesis is implemented in the $U(1)^{\prime}$ model. The numerical results for the predicted amount of BAU in the model are presented in Section~\ref{sec:numRes}. Section~\ref{sec:conclusion} concludes the paper.

\section{The Model}
\label{sec:model}
A generation-dependent $U(1)^{\prime}$ symmetry can play the role of a family symmetry which gives rise to realistic fermion mass hierarchy and mixing pattern through the Froggatt-Nielsen mechanism~\cite{ref:U1FModels,ref:SU5U1}. 
We will choose $U(1)^\prime$ charges such that the Lagrangian for the Yukawa interactions for all fermions, including the neutrinos, is given by
\begin{widetext}
\begin{equation}
\label{eqn:Lagrangian}
\mathcal{L}_{\mbox{\tiny Yukawa}} = Y^u Q\bar{u}H_1 + Y^d Q\bar{d}\bar{H}_2 + Y^e L\bar{e}\bar{H_2} + \frac{Y^{\nu}}{\Lambda} L\bar{\nu} H_1 \Xi + \frac{Y^{LLHH}}{\Lambda} LLH_1H_1 + Y^{\nu\nu} \bar{\nu} \bar{\nu} \;.
\end{equation}
\end{widetext}
Here $Y^{u}$, $Y^{d}$ and $Y^{e}$ are the effective Yukawa matrices for the up-type quark, down-type quark, and the charged lepton sectors; $Y^{\nu}$, $Y^{LLHH}$, and $Y^{\nu\nu}$ are  the Dirac, left-handed Majorana, and right-handed Majorana mass terms for the neutrinos, respectively. 

The higher-dimensional operator which generates effective Yukawa matrices in the Lagrangian are determined by the $U(1)^{\prime}$ charges of the SM fermions and that of  the Higgs, as
\begin{equation}
\label{eqn:Yukawa1}
Y_{\mbox{\tiny eff}}^{ij} = \biggl( Y_{ij} \frac{\phi}{\Lambda} \biggr)^{|q_i + q_j + q_H|} \, ,
\end{equation}
where $Y_{ij}$'s are $\sim \mathcal{O}(1)$ parameters in the renormalizable Lagrangian, with $i$ and $j$ being the generation indices; the parameters $q_i$ and $q_H$ are the $U(1)^{\prime}$ charges of the SM fermions and higgs field, respectively. The flavon fields $\phi$ and $\Xi$  
acquire vacuum expectation values once the $U(1)^{\prime}$ symmetry is broken, resulting in the effective Yukawa matrices as follows,
\begin{equation}
\label{eqn:Yukawa2}
Y_{\mbox{\tiny eff}}^{ij} = (Y_{ij} {\lambda})^{|q_i + q_j + q_H|} \,,
\end{equation}  
where $\lambda = \frac{\left< \phi \right>}{\Lambda} \sim 0.22$ is the Cabibbo angle. The hierarchical suppressions required among the Yukawa interactions can be obtained with proper choices of the $U(1)^{\prime}$ charges. 

Similarly, a natural small neutrino Dirac term can arise through similar higher-dimensional operators, 
\begin{equation}
\label{eqn:Yukawa2}
Y_{\nu}^{ij} = (Y_{ij} {\lambda})^{|q_i + q_j + q_H + q_{\Xi}|} \frac{\left<\Xi \right>}{\Lambda} \,,
\end{equation}  
with $\frac{\langle \Xi \rangle}{\Lambda} \sim \mathcal{O}(1)$. Furthermore, both the neutrino mass scale and flavor mixing are determined by the $U(1)^{\prime}$ charges. 

\begin{figure}[tb!]
\includegraphics[scale=0.9, angle = 0, width = 80mm, height = 25mm]{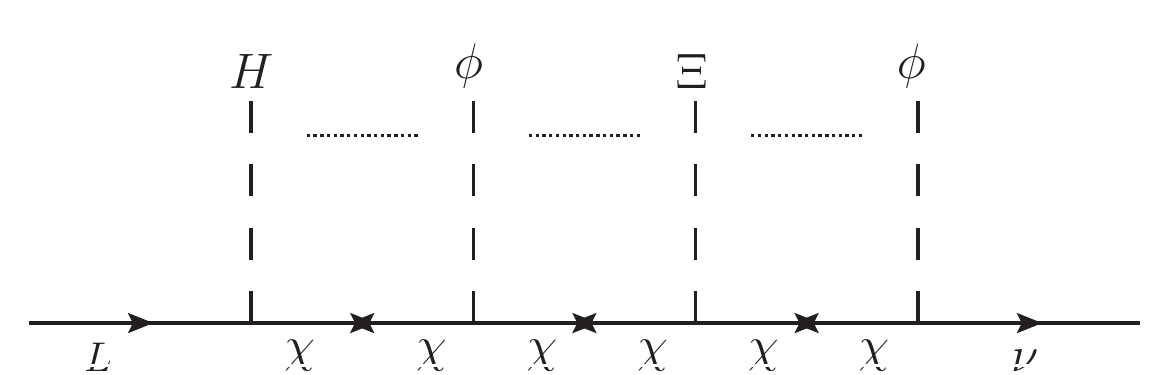}
	\caption{The effective Yukawa matrices are generated through higher dimentional operators. This is an schematic example for generating the Dirac mass term of the neutrinos. Here $H$ is the SM Higgs field and $\Xi \,, \phi$ are the flavon fields and $\chi$ fields are the vector-like fermions. Dotted lines between the flavons indicate that more $\phi$ fields are inserted as dictated by the $U(1)^{\prime}$ charge assignements. Upon the $U(1)^{\prime}$ breaking, flavons get VEV's and the heavy $\chi$ fields are integrated out, leading to the effective Yukawa matrices shown in Eq.~\ref{eqn:Yukawa2}.}
	\label{fig:FN}
\end{figure}

These operators can also be understood from the Feynman diagram shown in Fig.~\ref{fig:FN} where heavy vector-like fermions are involved. As we will discuss in Sec.~\ref{sec:diracLep}, these heavy fermions as well as the flavon field $\Xi$ appearing in the Dirac neutrino Yukawa interaction play a very important role for Dirac leptogenesis. 
The flavon field $\Xi$ and the scalar field $\xi$ which has opposite $U(1)^{\prime}$ charge of the $\Xi$ field are very important for generating the left- and right-handed lepton number asymmetries. These asymmetries are generated through their decays into the $\chi$ fields and the subsequent long decay chains of the $\chi$ fields to left- and right-handed leptons.

In our model, the $U(1)^{\prime}$ charges of the fermions are chosen in such a way that the fermions can be embedded into a $SU(5)$ GUT symmetry,
\begin{eqnarray}
q_{Q_i} & = & q_{\bar{u}_i} = q_{\bar{e}_i} \equiv q_{t_i} \; \\
q_{L_i} & = & q_{\bar{d}_i} \equiv q_{f_i} \; . 
\end{eqnarray}
Here $Q_{i}$, $u_{i}$, and $e_{i}$ are, respectively, the quark doublet, up-type quark singlet, and the charged lepton singlet, of the $i$-th generation, which are unifiable into a ${\bf 10}$ dimensional representation of $SU(5)$ and thus have the same $U(1)^{\prime}$ charge, $q_{t_{i}}$.  The lepton doublet, $L_{i}$, and the down-type quark singlet, $d_{i}$, of the $i$-th generation are unifiable into a ${\bf \bar{5}}$ of $SU(5)$ and thus they have the same $U(1)^{\prime}$ charge, $q_{f_{i}}$. 
The i-th generation right-handed neutrinos, $\bar{\nu}_i \;  (i = 1,2,3)$, carry the $U(1)^{\prime}$ charges $q_{N_i}$. All $U(1)^{\prime}$ charges are normalized with $q_{\phi} = -1$ where $q_{\phi}$ is the $U(1)^{\prime}$ charge of the flavon field, $\phi$. 

In terms of $q_{t_{i}}$ and $q_{f_{i}}$, the effective Yukawa matrices for the charged fermions can be simplified to 
\begin{equation}
Y_{ij}^u \sim (\lambda)^{|q_{t_i} + q_{t_j} + q_{H_1}|} \, ,
\end{equation}
\begin{equation}
Y_{ij}^d \sim (\lambda)^{|q_{t_i} + q_{f_j} - q_{H_2}|} \, ,
\end{equation}
and $Y^e = (Y^{d})^{T}$ because of the $SU(5)$ inspired charge choices. In the neutrino sector, the Dirac and the right-handed and left-handed Majorana mass terms are given by,
\begin{equation}
Y_{ij}^{\nu} \sim (\lambda)^{|q_{f_i} + q_{N_j} + q_{H_1} + q_\Xi|} \, ,
\end{equation}
\begin{equation}
Y_{ij}^{LLHH} \sim (\lambda)^{|q_{f_i} + q_{f_j} + 2q_{H_1}|} \, ,
\end{equation}
\begin{equation}
Y_{ij}^{\nu\nu} \sim (\lambda)^{|q_{N_i} + q_{N_j}|} \, .
\end{equation}

At this stage, there are 12 free parameters in total. The constraints from the anomaly cancellation conditions and the fermion mass hierarchy and mixings further reduce the number of free parameters. Given the simplified $U(1)^{\prime}$ charge assumption inspired by $SU(5)$ as shown above, the six anomaly cancellation conditions relating to the $U(1)^{\prime}$symmetry are reduced down to the following three~\cite{ref:SU5U1},
\begin{eqnarray}
[SU(5)]^{2}U(1)^{\prime}: & \sum_{i} (\frac{1}{2} q_{f_i} + \frac{3}{2} q_{t_i}) = 0 \; , \\
\mbox{gravity}-U(1)^{\prime}: & \sum_{i} (5 q_{f_i} + 10 q_{t_i} + q_{N_i}) = 0 \; , \\
U(1)^{\prime 3}: & \sum_{i} (5q_{f_i}^3 + 10 q_{t_i}^3 + q_{N_i}^3) = 0 \; .
\end{eqnarray}
In addition to the three anomaly cancellation conditions, there are 9 conditions from the fermion mass hierarchy and mixing constraints, and thus all of the free parameters are completely fixed.

To solve all the anomaly cancellation conditions is rather non-trivial. With the following parametrization, the $[SU(5)]^2U(1)^{\prime}$ and gravity-$U(1)^{\prime}$ conditions are automatically satisfied,
\begin{eqnarray} 
%\left\lbrace 
\begin{array}{lll} q_{t_1} & = & -\frac{1}{3} q_{f_1} - 2a \; , \nonumber \\
q_{t_2} & = & -\frac{1}{3} q_{f_2} + a + a^{\prime} \; ,  \nonumber \\
q_{t_3} & = & -\frac{1}{3} q_{f_3} + a - a^{\prime} \; ,  \end{array} \nonumber \,\,\qquad
%\left\lbrace 
\begin{array}{lll} q_{N_1} & = & -\frac{5}{3} q_{f_1} - 2b \; , \nonumber  \\
q_{N_2} & = & -\frac{5}{3} q_{f_2} + b + b^{\prime} \; , \nonumber \\
q_{N_3} & = & -\frac{5}{3} q_{f_3} + b - b^{\prime} \; . \end{array}
\label{eqn:param}
\end{eqnarray}
We then impose the following relations among the $U(1)^{\prime}$ charges which are well-motivated by the observed hierarchy and mixing patterns among the SM fermions: 
\begin{itemize}
\item[{(i)}] The fact that the third generation particles (top quark, bottom quark and tau) are heavy suggests that there is little or no suppression in their associated effective Yukawa couplings. Consequently it is natural to assume, 
\begin{equation}
|2q_{t_3} + q_{H_1}| = 0 \, , \quad |q_{t_3} + q_{f_3} - q_{H_2}| = 2 \, .
\end{equation}
\item[{(ii)}] To obtain the nearly tri-bimaximal mixing pattern in the neutrino sector, we require
\begin{eqnarray}
q_{f_2} = q_{f_3} \, , \quad |q_{f_1} - q_{f_2}| = 1 \, , \quad b^{\prime} = 0 \, ,\nonumber\\ 
q_{\nu_2} = q_{\nu_3} \, , \quad |q_{\nu_1} - q_{\nu_2}| = 1 \, .
\end{eqnarray}
\item[{(iii)}] By choosing $a = -7/9$ and $a^{\prime} = 1$, the resulting mass matrices give realistic quark masses and mixings which are consistent with experimental data. 
\end{itemize}

With the parametrization of the $U(1)^{\prime}$ charges and the assumed relations among the charges as motivated by the masses and mixing angles, the $[U(1)^{\prime}]^{3}$ anomaly cancellation condition is reduced to the following constraint on $q_{f_{2}}$ as a function of the charge splitting parameter $b$, as
\begin{equation}
q_{f_2} = \frac{-4550-2025b-2430b^2-729b^3}{45(124+90b+81b^2)} \; .
\end{equation}

Therefore, we have two free parameters left, which are $b$ and $q_{\Xi}$. The neutrino mass hierarchy can be  realized by choosing $b = -2/9$, which leads to $q_{f_2} = -13/15$ and $q_{\Xi} = 44/3$.  Both the left-handed and right-handed Majorana mass terms for the neutrinos are forbidden because all elements in these two mass matrices are non-integer powers of the expansion parameter, $\lambda$. As a result, neutrinos are predicted to be Dirac fermions with naturally suppressed masses. An alternative way to look at this is that the $U(1)^{\prime}$ breaks into a $Z_{3}$ discrete symmetry, under which the ${\bf \overline{5}}$ multiplets have charge $0$, and the ${\bf 10}$-plets have charge $1$, the right-handed neutrinos have charge $2$, and the Higgs fields have charge $1$. The $Z_{3}$ symmetry forbids the Majorana mass terms for the neutrinos~\footnote{We thank Michael Ratz for this observation.}.

We summarize the $U(1)^{\prime}$ charges in Table~\ref{tbl:u1Charge2}.  
\begin{table}[t!]
\begin{tabular}{c|c||c|c}\hline\hline\
Field & $U(1)^{\prime}$ charge & Field & $U(1)^{\prime}$ charge \\ \hline
$L_1, d_1$& $q_{f_1} = 2/15$ & $Q_1, u_1, e_1$ & $q_{t_1} = 68/45$ \\ \hline
$L_2, d_2$& $q_{f_2} = -13/15$ & $Q_2, u_2, e_2$ & $q_{t_2} = 23/45$ \\ \hline
$L_3, d_3$& $q_{f_3} = -13/15$ & $Q_3, u_3, e_3$ & $q_{t_3} = -67/45$\\ \hline   
$\nu_1$& $q_{N_1} = 2/9$ & $H_1$ & $q_{H_1} = 134/45$\\ \hline
$\nu_2$& $q_{N_2} = 11/9$ & $H_2$ & $q_{H_2} = -196/45$\\ \hline   
$\nu_3$& $q_{N_3} = 11/9$ & $\Xi$ & $q_{\Xi} = 44/3$\\ \hline  
$\phi$ & $q_{\phi} = -1$ & $\xi$ & $q_{\xi} = -44/3$\\ \hline
\end{tabular}
\caption{The $U(1)^{\prime}$ charges of the particles in the model.}   
\label{tbl:u1Charge2}
\end{table}
These charges lead to the following effective Yukawa matrix for the up-type quarks,
\begin{eqnarray}
\label{eqn:upYukawaNew}
Y^u & \sim & \left(\begin{array}{ccc} \lambda^{6} & \lambda^{5} & \lambda^{3} \\ \lambda^{5} & \lambda^{4} & \lambda^{2} \\ \lambda^{3} & \lambda^{2} & \lambda^{0} \end{array} \right) \; ,
\end{eqnarray}
and the effective Yukawa matrices for the down-type quarks and thus charged leptons are given by,
\begin{eqnarray}
\label{eqn:downYukawaNew}
Y^d & \sim & \left(\begin{array}{ccc} \lambda^{6} & \lambda^{5} & \lambda^{5} \\ \lambda^{5} & \lambda^{4} & \lambda^{4} \\ \lambda^{3} & \lambda^{2} & \lambda^{2} \end{array} \right) \; ,
\end{eqnarray}
\begin{equation}
Y^e \sim (Y^{d})^{T} \; .
\end{equation}
The effective neutrino Dirac mass matrix is given by, 
\begin{eqnarray}
Y^{\nu} & \sim & \left(\begin{array}{ccc} \lambda^{18} & \lambda^{19} & \lambda^{19} \\ \lambda^{17} & \lambda^{18} & \lambda^{18} \\ \lambda^{17} & \lambda^{18} & \lambda^{18} \end{array} \right) \; .
\end{eqnarray}
From this set of the $U(1)^{\prime}$ charge assignment, 6 vector-like fermions are needed for the up-type quark, down-type quark and charged lepton sector separately, and 19 for the neutrino sector, therefore, 37 in total. A vector-like fermion from a given sector carries the SM charges of the right-handed species it is granting mass to, so for instance a fermion in the charged lepton sector has only hypercharge, and one in the neutrino sector is a SM singlet. In the quark sector and charged lepton sector, their $U(1)^{\prime}$ charges differ by $-q_{\phi} = 1$ and the minimum of them related the up-type quark sector is $-202/45$ and those related the down-type quark and charged lepton sector is $-88/15$. It is more complicated for the neutrino sector because it is arbitrary to insert the $\Xi$ field in the effective operator $Y^{\nu}$. Here we use one particular choice as an example to illustrate the $U(1)^{\prime}$ charges of the vector-like fermions in the lepton sector, which is shown in the Fig.~\ref{fig:FN}, therefore, the minimum of the $U(1)^{\prime}$ is $-28/9$ and the rest of them will differ by $-q_{\phi} = 1$.

\section{Dirac Leptogenesis}
\label{sec:diracLep}

In general there are two distinct requirements for Dirac leptogenesis~\cite{ref:dirLep1} to work
in any given physics scenario. The first is that there be an adequate number
of out-of-equilibrium decays which have the appropriate CP violating behavior
to generate the needed asymmetry in left- versus right-handed lepton number.
The second requires that asymmetry to survive until beyond the time when the
sphaleron processes of $SU(2)_L$ decouple so that any asymmetry in baryon
number cannot be washed out by the same sphaleron processes which generated it.

There are a few different scenarios in which the first issue can be achieved. It
is possible that decays are significantly slower than annihilations, such that all
particles surviving annihilations are decaying out of equilibrium. It is also possible
that decays and inverse decays are the dominant processes which determine the number
density of the decaying state, and in that case only those decays happening after the
hubble parameter exceeds the decay rate will have the possibility of generating an
asymmetry. The decays can fall out of equilibrium at any temperature relative to the
mass of the field in question, depending on coupling strengths, and thus can lead to
late-decaying relics which are hot, warm, or cold~\cite{ref:KolbTurner}.

In our model the field which is decaying to produce a net left-handed lepton number
is the scalar $\Xi$, the lighter of the two scalars with $U(1)^\prime$ charge $|q|=44/3$. Any contribution from the heavier scalar $\xi$ will be equilibrated away by the interactions of $\Xi$, which freezes out at a lower temperature in the early universe. We assume that the additional flavon field $\phi$ is
light in comparison to these two, and that the heavier of these two is the dominant
source of $U(1)^{\prime}$ breaking at the epoch during which the decays are relevant, such that it becomes a real scalar boson through the Higgs mechanism. The relevant Lagrangian for the $\Xi$, $\xi$ fields is
\begin{equation}
\mathcal{L} \supset M_{\Xi}^2 |\Xi|^2 + M_{\xi}^2 |\xi|^2 + \lambda_1 \Xi \bar{\chi}_1 \chi_2 + \lambda_2 \xi \chi_1 \bar{\chi}_2 \; .
\end{equation}

In the case where annihilations are the dominant process changing the number density of the $\Xi$ fields, the freezeout temperature ($T_f$) and the variable $Y_{\Xi}$ (defined as $Y_{\Xi} \equiv n_{\Xi}/s$, where $n_{\Xi}$ is the number density of the $\Xi$ field while s is the entropy density, using the same convention as in~\cite{ref:KolbTurner}) are given by the well known relations,
\begin{equation}
Y_{\Xi} = \frac{3.79(n+1)x_f^{n+1}}{g_{*s}/\sqrt{g_{*}}M_{\mbox{\tiny{PL}}} M_{\Xi} \sigma_0} \; ,
\end{equation}
where $x_f$ is defined as $M_{\Xi}/T_f$ and is given by
\begin{widetext}
\begin{equation}
x_f = \ln\biggl[0.038(n+1)\frac{g}{\sqrt{g_{*}}}M_{PL}M_{\Xi} \sigma_0\biggr] - \biggl(n+\frac{1}{2}\biggr)\ln\biggl\{\ln\biggl[0.038(n+1)\frac{g}{\sqrt{g_{*}}}M_{\mbox{\tiny PL}}M_{\Xi}\sigma_0\biggr]\biggr\} \; ,
\end{equation}
In our model, at high temperatures relative to the vector-like fermion masses, we have $g = 1$ for our real boson $\Xi$ and $g_*= g_{*s}=327.75$. As for the washout effects, besides the inverse decay processes, there are also the annihilation processes which need to be taken into account. There are two distinct cases of interest for our model. The first is when $M_{Z^\prime}=g^\prime\langle\xi\rangle<M_\Xi$, such that annihilations to two vector bosons (shown in Fig.~\ref{fig:vvAnn}) are allowed. Note that $\langle\xi\rangle$ is controlled by the parameter $\lambda$ through the neutrino mass matrix.
\begin{figure}[tb!]
%\subfloat{\includegraphics[scale=0.65]{annil1.pdf} \label{fig:vvAnn-a}}
%\subfloat
{\includegraphics[scale=0.7]{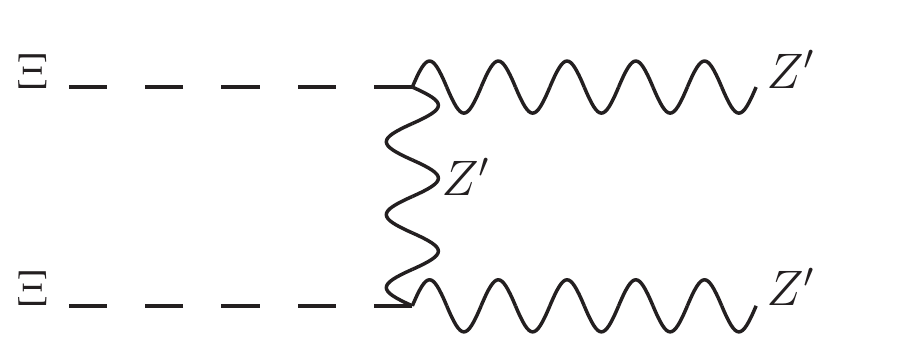} \label{fig:vvAnn-a}}
%\subfloat
{\includegraphics[scale=0.7]{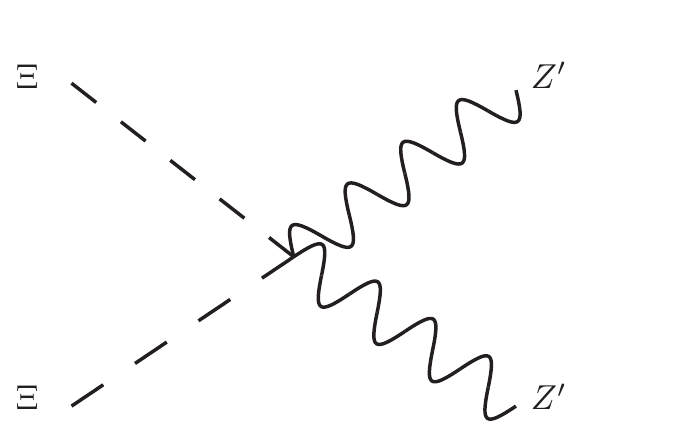} \label{fig:vvAnn-b}}
	\caption{Washout effects through annihilation processes into vector bosons.}
	\label{fig:vvAnn}
\end{figure}
In that case, there are s-wave annihilations corresponding to $n=0$ and 
%\begin{widetext}
\begin{equation}
%\sigma_0 = \biggl(\frac{44g'}{3\pi}\biggr)^4\frac{\sqrt{M_{\Xi}^2 - M_{Z'}^2}}{M_{\Xi}^3} \biggl[12\biggl(\frac{M_{Z'}}{M_{\Xi}}\biggr)^4 - 20\biggl(\frac{M_{Z'}}{M_{\Xi}}\biggr)^2 + 11\biggr].
\sigma_0 = \biggl(\frac{44g'}{3\pi}\biggr)^4\frac{\sqrt{M_{\Xi}^2 - M_{Z'}^2}}{8\pi M_{\Xi}^3} \biggl[4\biggl(\frac{M_{Z'}}{M_{\Xi}}\biggr)^8 + 4\biggl(\frac{M_{Z'}}{M_{\Xi}}\biggr)^6-\biggl(\frac{M_{Z'}}{M_{\Xi}}\biggr)^4 + 2\biggl(\frac{M_{Z'}}{M_{\Xi}}\biggr)^2 + 3\biggr] .
\end{equation} 
%\end{widetext}
If annihilations to vector bosons are kinematically forbidden then the dominant annihilation channel is to two vector-like fermions through yukawa couplings (shown in Fig.~\ref{fig:ffAnn}).
\begin{figure}[tb!]
\includegraphics[scale=0.7]{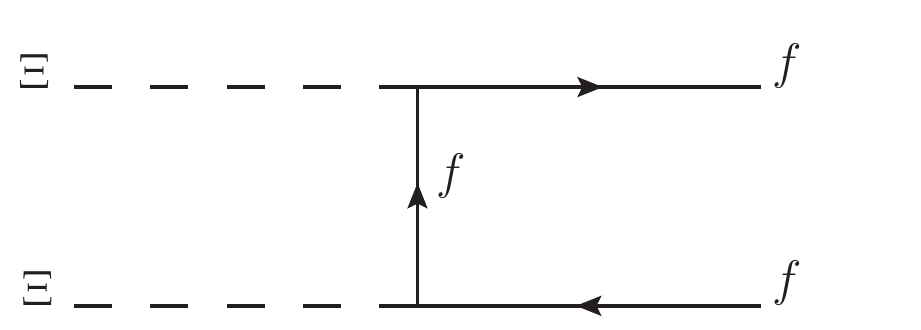}
\caption{Washout effects through annihilation into fermion pairs.}
\label{fig:ffAnn}
\end{figure}

The annihilation cross section, assuming identical $\chi$ masses, is,
%\begin{widetext}
\begin{equation}
%\sigma_0= \frac{6|\lambda_1|^4}{32\pi} \frac{\sqrt{M_{\Xi}^2 - m_{\chi}^2}}{M_{\Xi}^7}\biggl[2M_{\Xi}^4 - 44M_{\Xi}^2m_{\chi}^2 + \frac{352}{3}m_{\chi}^4 - \frac{208}{3}\biggl(\frac{m_{\chi}^6}{M_{\Xi}^2}\biggr)\biggr] \; \; (n = 1) \; .
\sigma_0 = \frac{\lambda^4}{\pi} \frac{m_{\chi}^2(M_{\Xi}^2 - m_{\chi}^2)^{3/2}}{M_{\Xi}^7} \; .
\end{equation}
\end{widetext}
In the case where decays are dominant, the freezeout temperature is identified by simply requiring
\begin{equation}
\label{eqn:fzcond}
\frac{\Gamma}{2H(T)} \biggr \vert_{T = Tf} = 1,
\end{equation}
in which we follow the convention in~\cite{ref:KolbTurner}, where $\Gamma$ is the interaction rate per particle for the reaction that keeps the species in thermal equilibrium and H is the expansion rate of the Universe. The out-of-equilibrium abundance is taken to be the equilibrium abundance at the time of decay decoupling. We consider three distinct regimes for this, such that
\begin{equation}
Y_\Xi=\begin{cases}
0.278 \frac{g}{g_{*s}} \quad (x_f\leq1) \\
0.278\frac{g}{g_{*s}}\biggl[2\left(\left(\frac{3}{2}\right)^{3/2}e^{-3/2} - 1\right)\left(x_f-1 \right) + 1\biggr ] \\
\quad \quad \quad \quad \quad (1 < x_f < 1.5) \\
0.278 \frac{g}{g_{*s}}x_f^{3/2}e^{-x_f} \quad (x_f \geq 1.5) 
\end{cases} \; .
\end{equation}
The central region with $1 < x_f < 1.5$ is chosen as a simple interpolation between the two well-known limiting cases of hot and cold equilibrium conditions.

\begin{figure}[tb!]
\includegraphics[scale=0.4, angle = 0, width = 50mm, height = 30mm]{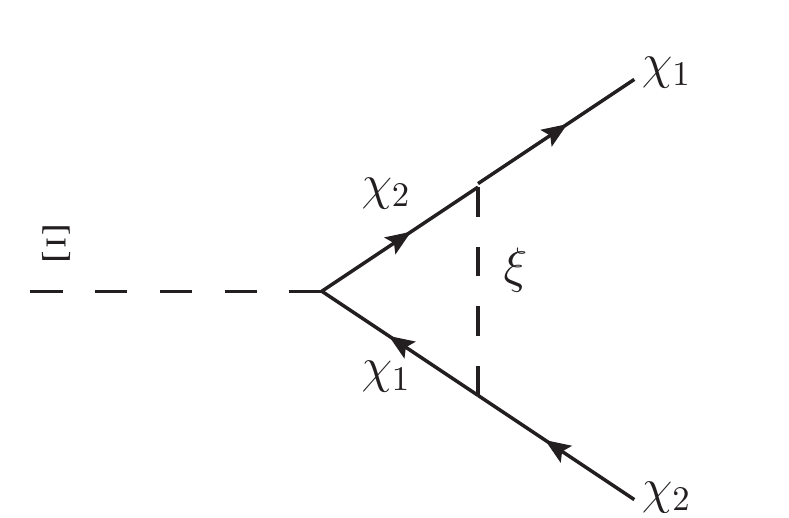}
	\caption{The vertex correction diagram contributing to the CP violation.}
	\label{fig:tri-ang}
\end{figure}

\begin{figure}[tb!]
\includegraphics[scale=0.4, angle = 0, width = 80mm, height = 30mm]{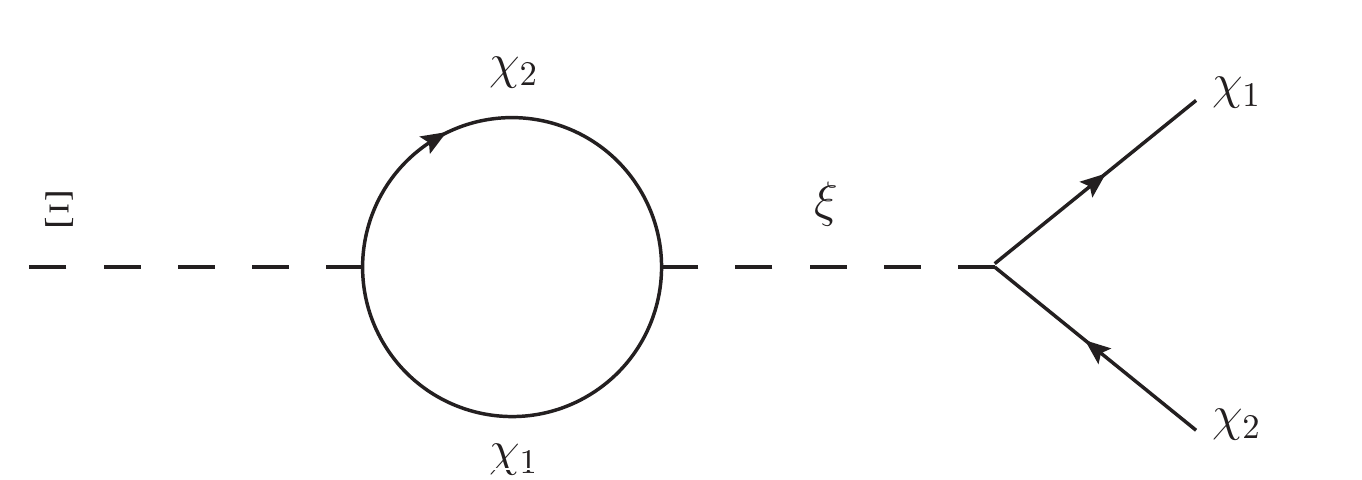}
	\caption{The self-energy loop diagram contributing to the CP violation.}
	\label{fig:self-eng}
\end{figure}

Once we have the number of states which will be decaying out of equilibrium, we still need to know how much they favor left-handed leptons over anti-leptons. This is encoded in the variable
\begin{equation}
\epsilon_\Xi = \mbox{BR} \left(\Xi\to\chi_1\bar{\chi_2}\right)-\mbox{BR} \left(\Xi\to\chi_2\bar{\chi_1}\right) \; .
\end{equation}
This gives rise to an asymmetry in left and right handed lepton number because $\chi_1$ cascade decays to give a left-handed lepton doublet, while $\chi_2$ decays to a right-handed neutrino. The entire process is a particular crossing of the Feynman diagram in FIG.~\ref{fig:FN} such that the initial state particle is the $\Xi$ and all other fields are in the final state. In our model this quantity is given
by
\begin{equation}
\epsilon_{\Xi} = \frac{8 \mbox{Im} [\lambda_1^2 \lambda_2^2] \mbox{Im} [I_{\Xi\xi}]}{\Gamma_{\Xi}} \; ,
\end{equation}
where total decay width of $\Gamma_{\Xi}$ has two parts which are
\begin{equation}
\Gamma_{\Xi \rightarrow \chi1 \chi2} = \frac{|\lambda_1|^2 \sqrt{M_{\Xi}^2 - 4m_{\chi}^2}(M_{\Xi}^2/2 - m_{\chi}^2)}{16\pi M_{\Xi}^2} \; ,
\end{equation}
and
\begin{equation}
\Gamma_{\Xi \rightarrow Z^{\prime} Z^{\prime}} = \frac{4(44/3)^2 M_{Z'}^2 \sqrt{M_{\Xi}^2 - 4M_{Z'}^2}}{16\pi M_{\Xi}^2} \; ,
\end{equation}
which only contributes if $M_\Xi>2M_{Z^{\prime}}$. In our case the kinematic contributions to the CP asymmetry are from interferences between the tree level diagram and the vertex correction  (Fig.~\ref{fig:tri-ang}) and self-energy loop diagrams (Fig.~\ref{fig:self-eng}) in the $\Xi$ field decaying into fermion $\chi_1 \chi_2$ channel~\cite{Liu:1993j}.
\begin{equation}
\mbox{Im} [I_{\Xi\xi}] = \frac{M_{\Xi}\biggl(1 - \left( \frac{M_{\xi}}{M_{\Xi}} \right)^{2} \mbox{ln} 
\left[ 1 + \left( \frac{M_{\Xi}}{M_{\xi}} \right)^2 \right] - \frac{M_{\Xi}^2}{M_{\Xi}^2 - M_{\xi}^2}\biggr)}{32\pi^{2}} \; .
\end{equation}
%We also need to take the contribution from the $\xi$ field into account by calculating the %variables $\epsilon_{\xi} = \mbox{BR} \left(\xi\to\bar{\chi_1}\chi_2\right)-\mbox{BR} %\left(\xi\to\bar{\chi_2}\chi_1\right)$ and $Y_{\xi} \equiv n_{\xi}/s$ (anologous to the %$\epsilon_{\Xi}$ and $Y_{\Xi}$ of the $\Xi$ field) and it can be done by simply interchanging the %massess of them ($m_{\Xi} \leftrightarrow m_{\xi}$). 
Ultimately, the amount of BAU generated is equal to~\cite{ref:dirLep1},
\begin{equation}
\eta_B = -\frac{28}{79}n_{\nu_R} = -\frac{28}{79}\epsilon_{\Xi}Y_{\Xi}\; .
\end{equation}

The connection between the low energy scale parameters such as the SM fermion masses and mixing angles and the parameters related to the UV complete theory (e.g, $\lambda_1,\; \lambda_2,\; M_{\Xi},\; M_{\xi}, m_{\chi}$) is fairly weak since the effective Yukawa interactions are generated through higher dimensional operators and the effects from the UV complete theory has been integrated out and thus are indirect. While the UV physics does, ultimately, give rise to the masses and mixings of the SM fields, many parameters enter in to generate those masses and mixings, so measuring the masses and mixings doesn't tell us much about individual parameters of the UV theory which is responsible, in particular, for leptogenesis. Thus, there is not a strong prediction from the success of leptogenesis for low-scale parameters in this theory.

\section{Numerical Results}
\label{sec:numRes}

Given the analysis above, we have in total nine free parameters. These are $M_{\Xi}$, $M_{\xi}$, $m_{\chi1}$, $m_{\chi2}$, Re[$\lambda_1$], Im[$\lambda_1$], Re[$\lambda_2$], Im[$\lambda_2$], and $\alpha$%. Among these parameters, $\lambda_1$ and $\lambda_2$ are complex 
, where $\alpha = g'^2/4\pi$. Given the values of $\alpha$ and $\left< \xi \right>$ (which varies as we search the allowed parameter space), the value of $M_{Z^{\prime}}$ is determined. We first make the simplification,  $\lambda_1 = \lambda_2 = |\lambda_1| e^{-i \pi/8}$, which leads to the maximum imaginary phase in $\mbox{Im} [\lambda_1^2 \lambda_2^2]$ in the CP violation factor. In addition, we take the masses of fermions to be  $m_{\chi1} = m_{\chi2} = 0.4\, M_{\Xi}$. 
We then search for the allowed parameter space that gives rise to the baryon number asymmetry within the $1\, \sigma$ allowed range, $\eta_B = (6.19 \pm 0.15) \times 10^{-10}$~\cite{ref:barAsyExp}. First, we investigate how the couplings $\lambda_{1}$ and $\alpha$  affect the baryon number asymmetry. For the fixed hierarchical masses of $M_{\Xi} = 5.7 \times 10^{6}$ GeV and $M_{\xi} = 4.9\, M_{\Xi}$, we find that there is no constraint on $\alpha$ up to the perturbativity bound. For $|\lambda_1|$ in the small band,  $0.3925 \leq |\lambda_1| \leq 0.4063$, the baryon number asymmetry within the $1 \sigma$ allowed range can be generated. 
%In the presence of additional washout effects other than those discussed in Sec.~\ref{sec:diracLep}, successful Dirac leptogenesis can still be possible in the model and 
To have a sufficient BAU, $\eta_B \geq 6.19 \times 10^{-10}$, an upper limit $|\lambda_1| \leq 0.4$ is obtained. This is because, in the low $\Xi$ mass region, the dominant wash-out process is the annihilation into fermions which is proportional to $\lambda_1^4$, so larger $\lambda_1$ leaves too small a frezeout relic of $\Xi$. 
%The lower limit on $|\lambda_1|$ arises due to the requirement of having sufficient CP violation.

%In addition, we also investigate the dependence of the asymmetry on the $U(1)^{\prime}$ breaking scale as well as the mass ratio between $\Xi$ and $\xi$ fields. We fix the couplings $|\lambda_1| = |\lambda_2| = 0.4$, and $\alpha = 0.8$. With additional washout effects, there are still large regions of parameter space which give rise to the enough baryon number asymmetry. This is demonstrated in the $\mbox{Log}_{10} (M_{\xi}/M_{\Xi}) \sim \mbox{Log}_{10} (M_{\Xi})$ figure (Fig.~\ref{fig:mRatioMAll}).  
\begin{figure}[bt!]
\includegraphics[scale=0.9, angle = 0, width = 90mm, height = 80mm]{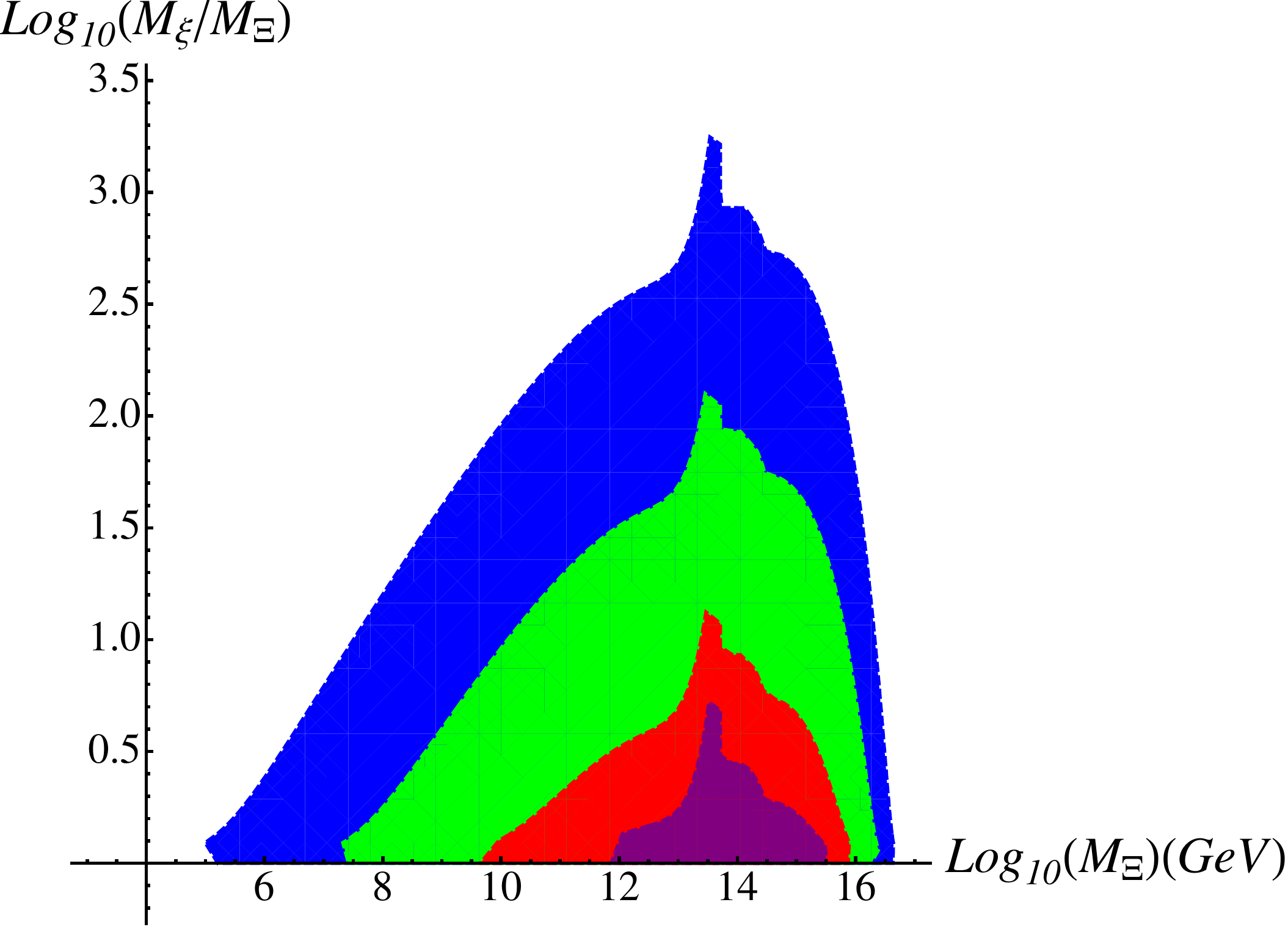}
	\caption{The allowed regions on the $\mbox{Log}_{10} (M_{\xi}/M_{\Xi}) - \mbox{Log}_{10} (M_{\Xi})$ plane  with model parameters $|\lambda_1| = |\lambda_2| = 0.4$, $\alpha = 0.8$ and $m_{\chi1} = m_{\chi2} = 0.4 m_{\Xi}$. The purple region corresponds the baryon number asymmetry $\eta_B \geq 6.19 \times 10^{-5}$. The red region corresponds to $\eta_B \geq 6.19 \times 10^{-6}$, the green region corresponds to $\eta_B \geq 6.19 \times 10^{-8}$, and the blue region corresponds to $\eta_B \geq 6.19 \times 10^{-10}$.}
	\label{fig:mRatioMAll}
\end{figure}

In addition, we also investigate the dependence of the asymmetry on the $U(1)^{\prime}$ breaking scale as well as the mass ratio between $\Xi$ and $\xi$ fields. This is presented in Fig.~\ref{fig:mRatioMAll} where regions of the parameter space on the $M_{\Xi}$ -- $M_{\xi}/M_{\Xi}$ plane corresponding to different amounts of the asymmetry is shown. In the figure, the couplings are taken to be $|\lambda_1| = |\lambda_2| = 0.4$, and $\alpha = 0.8$, as a result, $M_{Z^{\prime}} = 46.5 M_{\Xi} > M_{\Xi}$ and the decay and annihilation of $\Xi$ through the $Z^{\prime}$ boson shut off. The regions of different colors correspond to the allowed regions with different lower bounds of the baryon number asymmetry.  The allowed parameter space exhibits the following features. First of all, the masses of the $\Xi$ and $\xi$ fields can be as light as $100$ TeV. In other words, sufficient amount of baryon number asymmetry can be generated even if the $U(1)^{\prime}$ symmetry breaking scale is as low as $\sim 100$ TeV which naturally satisfies the constraints from FCNCs.
%The sharp cut-off on $M_{\Xi}$ at around 100 TeV is due to the constraints from FCNC's and there is little constraint from the requirement of generating sufficient BAU. 
Furthermore, the baryon number asymmetry can be enhanced to be as large as $\mathcal{O}(1)$ due to the presence of the resonance effects through the self-energy diagram, though the region of parameter space where this occurs is too small to visualize in Figure \ref{fig:mRatioMAll}. In addition, in the region where the value of $M_{\Xi}$ is smaller than around $\mathcal{O}(10^{14})$ GeV, the asymmetry increases as $M_{\Xi}$ increases. This is because the dominant wash-out process is annihilations and the cross section is inversely proportional to $M_{\Xi}^2$. As a result, the BAU is roughly proportional to the mass of $\Xi$. On the other hand, the two kinks for each colored region where $M_{\Xi}$ is greater than around $10^{14}$ GeV is due to the interpolation from $x_{f} < 1$ into $1 < x_{f} < 1.5$ and from $1 < x_{f} < 1.5$ into $x_{f} > 1.5$. We have confirmed this numerically.  
 
\section{Conclusion}
\label{sec:conclusion} 

We propose a model for Dirac leptogenesis based on a non-anomalous $U(1)^{\prime}$ gauged family symmetry. The anomaly cancellation conditions, including the $[U(1)^{\prime}]^{3}$ condition, are satisfied with no new chiral fermions other than three right-handed neutrinos, giving rise to stringent constraints among the $U(1)^{\prime}$ charges of the fermions. 

Realistic masses and mixing angles are obtained for all fermions, including all quarks and leptons. The model predicts neutrinos of the Dirac type with naturally suppressed masses. Dirac leptogenesis is achieved through the decay of the flavon fields. The cascade decays of the vector-like heavy fermions in the Froggatt-Nielsen mechanism play a crucial role in the separation of the primordial lepton numbers. 

We identify the annihilation processes that are present in the model. We then perform a detailed study of the allowed parameter space in the model. We find that a large region of the parameter space gives rise to sufficient amount of the cosmological baryon number asymmetry through Dirac leptogenesis.

\begin{acknowledgments}
We would like to thank Michael Ratz for his invaluable comments. The work of MCC and JH was supported, in part, by the National Science Foundation under Grant No. PHY-0970173. The work of WS is supported in part by the NSF under grant PHY-0970171. This material is also based upon work supported in part by the National Science Foundation under Grant No. 1066293 and the hospitality of the Aspen Center for Physics. WS and JH thank the hospitality of TASI, where part of the work was finished.  
\end{acknowledgments}

\end{document}